\documentclass[mathleft,fleqn,%
]{an}
\usepackage{graphicx}
\usepackage[varg]{txfonts}
\begin{document}

% The following seven commands are intended for editorial usage and
% should be ignored by the author(s).
\Pagespan{1}{}% Document's page range. 
% If second parameter is left empty, the last page is computed
% automatically.
\Yearpublication{2017}%
\Yearsubmission{2016}%
\Month{0}%   
\Volume{999}%  
\Issue{0}% 
\DOI{asna.201400000}% 

%%%%%%%%%%%%%%%%%%%%%%%%%%%%%%

\title{A TGAS/{\em Gaia} DR1 parallactic distance to the $\sigma$~Orionis cluster}

\author{Jos\'e A. Caballero\inst{1,2}\fnmsep\thanks{\tt caballero@cab.inta-csic.es}
%\and  G.\,H. Ostwriter\inst{2,3}
}
\titlerunning{A TGAS parallactic distance to $\sigma$~Orionis}
\authorrunning{J.\,A. Caballero}
\institute{
Landessternwarte K\"onigstuhl, Zentrum f\"ur Astronomie der Universit\"at Heidelberg, K\"onigstuhl 17, D-69117 Heidelberg, Germany
\and  {Centro de Astrobiolog\'{\i}a (CSIC-INTA), ESAC campus, Camino Bajo del Castillo s/n, E-28692 Villanueva de la Ca\~nada, Madrid, Spain}
}

\received{2016 Nov 19}
\accepted{2017 February 20}
\publonline{2017 April dd}


\abstract{%
With the new Tycho-{\em Gaia} Astrometric Solution, I~derive a new parallactic distance to the young $\sigma$~Orionis open cluster, which is a cornerstone region for studying the formation and evolution of stars and substellar objects from tens of solar masses to a few Jupiter masses.
I~started with the list of the 46 brightest cluster stars of Caballero (2007).
After identifying the 24 TGAS stars in the 30\,arcmin-radius survey area and accounting for 11 FGKM-type dwarfs and giants in the fore- and background, I~got a list of 13 cluster members and candidates with new parallaxes.
Of them, I~discarded five cluster member candidates with questionable features of youth and/or discordant parallaxes and proper motions, including a distant Herbig Ae/Be star, and proceeded with the remaining eight stars.
The $\sigma$~Orionis mean heliocentric distance is $d =$ {360}$^{+44}_{-35}$\,pc, which is consistent with a contemporaneous interferometric determination that concludes a two-decade dispute on the cluster distance. 
As a by-product, the re-classification of those five cluster member candidates, now interlopers, shows a manifest deficit of cluster stars between 1.2 and 2.1\,$M_\odot$, which leaves the door open to new astrometric membership analyses with the next {\em Gaia} data releases.
}

\maketitle

%------------------------------------------------------------------------------

\section{Introduction}
\label{section.introduction}

The \object{$\sigma$~Orionis} cluster (Garrison 1967; Wolk 1996; Sherry et~al. 2004; Walter et~al. 2008; Caballero 2008c) is located in the easternmost part of the \object{Ori~OB1b} association and is one of the most attractive and visited regions for night sky observers. 
The cluster gets the name from the massive star system in its centre, which is the fourth brightest star in the Orion Belt and visible with naked eye.
The $\sigma$~Orionis cluster is important for several reasons:
\begin{itemize}
\item Its stars illuminate the Horsehead Nebula photodissociation region (Pound et~al. 2003; Pety et~al. 2005, 2012; Goicoechea et~al. 2006).
\item It contains an abundant X-ray emitter population (Mokler \& Stelzer 2002; Franciosini et~al. 2006; Skinner et~al. 2008; Caballero et~al. 2009, 2010).
\item It is a cornerstone for studying discs and their frequency at an age of about 3\,Ma and at all mass domains (Oliveira et~al. 2004; Caballero et~al. 2007; Zapatero O\-so\-rio et~al. 2007; Hern\'andez et~al. 2007, 2014; Luhman et~al. 2008; Williams et~al. 2013; Mauc\'o et~al. 2016).
\item The prototypical, helium-rich, magnetically active, B2\,Vp star $\sigma$~Ori\,E is in its centre (Walborn 1974; Groote \& Hunger 1982; Townsend et~al. 2005, 2010; Oksala et~al. 2012; Caballero et~al. 2016).
\item It holds four Herbig-Haro objects and dozens Herbig Ae/Be and T~Tauri stars and brown dwarfs (Haro \& Moreno 1953; Reipurth et~al. 1998; Andrews et~al. 2004; Caballero et~al. 2006; Rigliaco et~al. 2011b; Riaz et~al. 2015).
\item The central trapezium-like star system contains the most massive ``binary'' with an astrometric orbit (Hartkopf et~al. 1996; Mason et~al. 1998; Caballero 2008b; Sim\'on-D\'{\i}az et~al. 2011, 2015; Schaefer et~al. 2016).
\item Its proximity, youth, and low extinction facilitate studies of photometric variability, accretion rates, and accretion frequency at low and very low masses (Zapatero Osorio et~al. 2002; Caballero et~al. 2004; Scholz \& Eisl\"offel 2004; Kenyon et~al. 2005; Gatti et~al. 2008; Sacco et~al. 2008; Scholz et~al. 2009; Cody \& Hillenbrand 2010; Rigliaco et~al. 2011a).
\item It is the star-forming region with the largest number of confirmed brown dwarfs ($\mathcal{M} \lesssim$ 72\,$M_{\rm Jup}$) and objects below the deuterium burning mass limit (i.e., isolated planetary-mass objects; $\mathcal{M} \lesssim$ 13\,$M_{\rm Jup}$) with spectroscopy and youth features (B\'ejar et~al. 1999, 2011; Zapatero Osorio et~al. 2000; Caballero et~al. 2007, 2012; Bihain et~al. 2009; Lodieu et~al. 2009; Rigliaco et~al. 2012; Pe\~na-Ram\'{\i}rez et~al. 2012).
\end{itemize}

The abundance of known substellar objects has led $\sigma$~Orionis to become a key region all over the sky to study the initial mass function, especially at very low masses.
However, one of the main caveats in the initial mass function determination in open clusters in general, and in $\sigma$~Orionis in particular, is the uncertainty in heliocentric distance.
Its contribution to the final errors is comparable to the uncertainty in age, and far larger than those in theoretical models, contamination rate, or multiplicity (Caballero 2011).
Similarly, a more precise and accurate distance to $\sigma$~Orionis would help in improving the determination of disc life-times, X-ray luminosities, and the three-dimensional structure of the Ori~OB1b association, among other parameters.  

There have been different determinations of the distance to the Ori~OB1b association and $\sigma$~Orionis cluster, which vary between 352$^{+166}_{-168}$\,pc and 473$\pm$33\,pc (see below).
The closer the cluster, the smaller the distance modulus and the lower the member luminosities.
A change of 25\,\% in the assumed distance can have a significant effect on the derived individual masses, star-to-brown dwarf ratio, or slope and minimum-mass cut-off of the initial mass function, just to cite a few examples (Lodieu et~al. 2009).

With the aim of improving previous determinations and of comparing with the precise, contemporaneous, interferometric determination of a distance by Schaefer et~al. (2016), I~take advantage of the recent release of the Tycho-{\em Gaia} Astrometric Solution (TGAS; {\em Gaia} Collaboration et~al. 2016) for narrowing down a new parallactic distance to $\sigma$~Orionis.

\section{Analysis and results}
\label{section.analysis+results}

%__________________________________________________ 
   \begin{table}
      \caption[]{Youth features of the {13} Mayrit/TGAS cluster member candidates.} 
         \label{table.mayrit}
     $$ 
         \begin{tabular}{lll}
            \hline
            \hline
            \noalign{\smallskip}
Mayrit 		& Youth feature			& Reference$^{a}$ \\
            \noalign{\smallskip}
            \hline
            \noalign{\smallskip}
182305  	& OB spectral type  			& SC71	\\ % HD 294272 A
		& X-rays					& Cab07, Cab10 \\
189303  	& OB spectral type   			& SC71, Gue81	\\ % HD 294272 B        
		& X-rays					& Fra06, Cab07, Cab10 \\
(459340)  	& H$\alpha$			   	& Ste86, DK88, Cab08	\\ % StHA 50: A2-6Ve, -10+-1 AA
		& Debris disc				& Cab08, Her14  \\ % SO595 Her14 DD/EV Table 8   
524060  	& X-rays					& Cab07, Cab10	\\ % HD 37564: A8V Gue81, A6V Her14 Binary candidate
		& Debris disc				& TO91, Her07, Mit15	\\
(717307)	& X-rays?					& Fra06	\\ % [W96] 4771-950, SO210457, NX 6
		& Li~{\sc i}?				& Cab06	\\ %pEW(Li I) = +0.07+-0.01 AA
783254	& Li~{\sc i}				& Wol96	\\ % 2E 1455, Wol96, Cab12: K0:, Halpha abs. 
		& X-rays, flare				& Fra06, Cab07, Cab09, Cab10	\\ % Violent long-lived flare		
863116	&  Li~{\sc i}, H$\alpha$		& Alc96	\\ % RX = HD 294300
		& X-rays, flare				& Fra06, Cab09, Cab10	\\ 
960106	& OB spectral type			& JB81	\\ % V1147 Ori
		& X-rays					& Cab10 \\ 
		& Phot. variability			& Nor84	\\
1160190	& None					& ...	\\ % HD 294279, SO561 Her14
(1227243)& None					& ...	\\ % HD 294275, SO82 Her14
1285339	& Li~{\sc i}, H$\alpha$		& Tor95, Roj08	\\ % HD 294268, G3V
		& Transition disc			& GL90, OvL04, Her07, Wil13	\\
		& X-rays					& Cab07	\\
(1456284)& X-rays?					& LSC08	\\ % TYC 4770-1261-1
(1468100)& Fast rotation?			& GC93	\\ % HD 294301
\noalign{\smallskip}
            \hline
         \end{tabular}
     $$ 
\begin{list}{}{}
\item[$^{a}$] {\em References.}
SC71: Schild \& Chaffee 1971;
Gue81: Guetter 1981;
JB81: Joncas \& Borra 1981;
Nor84: North 1984;
Ste86; Stephenson 1986;
DK88: Downes \& Keyes 1988;
GL90: Garc\'{\i}a-Lario et~al. 1990;
TO91: Tovmasyan \& Oganesyan 1991;
GC93: Gray \& Corbally 1993;
Tor95: Torres et~al. 1995;
Alc96: Alcal\'a et~al. 1996;
Wol96: Wolk 1996;
OvL04: Oliveira \& van~Loon 2004;
Fra06: Franciosini et~al. 2006;
Her07: Hern\'andez et~al. 2007;
Cab08: Caballero et~al. 2008; % ALFOSC
LSC08: L\'opez-Santiago \& Caballero 2008;
Roj08: Rojas et~al. 2008;
Cab09: Caballero et~al. 2009; % ROSAT
Wil13: Williams et~al. 2013;
Her14: Hern\'andez et~al. 2014;
Mit15: Mittal et~al. 2015.
\end{list}
   \end{table}

First, I retrieved the list of the 46 brightest stars in the $\sigma$~Orionis cluster compiled by Caballero (2007).
They are the 41 stars from the Tycho-2 catalogue (TYC2; H{\o}g et~al. 2000) within a 30\,arcmin-radius area centred on the  \object{$\sigma$~Ori\,A}a,Ab,B triple star (Sim\'on-D\'{\i}az et~al. 2015), plus the remaining four Tycho-1 (TYC1; H{\o}g et~al. 1998) stars in the area not tabulated by TYC2, and $\sigma$~Ori\,C, which was skipped by {\em Hipparcos} in spite of its relative brightness ($V \approx$ 8.8\,mag) and which proper motion was taken from the NOMAD catalogue (Zacharias et~al. 2004).
The radius choice was a~posteriori ratified by Caballero (2008a¶), who determined that the $\sigma$~Orionis cluster is composed by a dense core that extends from the centre to $\sim$20\,arcmin and a rarified halo at larger separations.
At more than 30\,arcmin from the centre, the contamination by neighbouring young populations at similar heliocentric distances gets significant: 
the younger \object{Alnitak} and \object{Horesehead Nebula} populations to the north-east and east (Caballero 2008a),
the outskirts of the \object{Orion Nebula Cluster} to the south and south-west,
and the slightly older cluster around \object{Alnilam} to the north-west (Caballero \& Solano 2008; Kubiak et~al. 2016). 
The 46 bright stars are sorted in Table~\ref{table.brightest} by angular separation to the cluster centre.

In the circular area, {\em Hipparcos} tabulated parallaxes for only five stars (actually four, since $\sigma$~Ori\,D was given the parallax of  $\sigma$~Ori\,Aa,Ab,B).
Because of occasional saturation and mixing of the signals of the three components (Schaefer et~al. 2016), the new {\em Hipparcos} data reduction of van~Leeuven (2007) provided larger uncertainties and less reliable values for the parallax of $\sigma$~Ori\,Aa,Ab,B than the original reduction by Perryman et~al. (1997), so I kept the original values for homogeneity (denoted by HIP1 in last column of Table~\ref{table.brightest}).
Three of the four independent values clustered at around $\pi \approx$ 2.5--2.8\,mas ($d \approx$ 350--400\,pc).

Of the 46 stars in Table~\ref{table.brightest}, 39 were listed in the {\em Gaia} first data release (DR1; {\em Gaia} Collaboration et~al. 2016).
Five of the seven missing stars are the brightest and most massive ones towards the cluster: $\sigma$~Ori\,Aa,Ab,B (unresolved), D, and~E, \object{HD~37699}, and \object{HD~294271} ($V_T \approx$ 3.8--7.9\,mag, $\mathcal M \approx$ 3.9--17\,$M_\odot$).
The other two stars, \object{HD~37686} and \object{TYC~4771--934--1}, have expected $G$ magnitudes of about 9.2\,mag and 11\,mag and are less affected than other fainter {\em Gaia} DR1 stars by the bright emission nebula \object{IC~434}, against which the Horsehead Nebula is silhouetted.
In general, the {\em Gaia} $G$-band mean magnitude smoothly interpolates the Tycho $B_T$, $V_T$ (H{\o}g et~al. 2000), 2MASS  $J$, $H$, $K_s$ (Skrutskie et~al. 2006), and AllWISE $W1$, $W2$, $W3$, and $W4$ (Cutri et~al. 2014) magnitudes.

Twenty-four of the 39 bright {\em Gaia} DR1 stars in Table~\ref{table.brightest} are tabulated in TGAS and have new parallax determinations.
As expected, there is no other TGAS star in the surveyed area.
This list of {24} stars includes {13} of the 29 bright, resolved, young stars and candidates in the $\sigma$~Orionis cluster identified by Caballero (2007).
The {13} stars have a Mayrit designation in the second column of Table~\ref{table.brightest}, and a TGAS reference in the last one.
The Mayrit catalogue tabulated over 300 $\sigma$~Orionis members and member candidates identified in a 2MASS+DENIS search accompanied by an exhaustive literature compilation of youth features (Caballero 2008c). 

The other {11} TGAS stars in the 30\,arcmin-radius area are either fore- or background interlopers that contaminate photometric and astrometric searches in the cluster.
The new TGAS data reinforce this classification, based on the stars position in colour-magnitude diagrammes, parallactic distances (from 193$\pm$30\,pc to 1900$\pm$1000\,pc) or total proper motions (from 4.0\,mas\,a$^{-1}$ to 44.3\,mas\,a$^{-1}$), apart from their known spectra in some cases.
The poorly investigated stars \object{TYC~4770--873--1} and \object{TYC~4770--1129--1} were photometric cluster member candidates in Caballero (2007), but later were not listed as candidate members in the Mayrit catalogue.
The 11 stars seem to be late-F/early-G dwarfs and K/M giants. % (Table~\ref{table.brightest}).

I investigated in detail the actual cluster membership of the {13} Mayrit/TGAS stars, all of which were expected to be members in the young $\sigma$~Orionis cluster.
However, since the publication of the Mayrit catalogue by Caballero (2008c), there have been numerous works devoted to study stars in the cluster with a variety of techniques (optical spectroscopy, X-rays, proper motion, mid-infrared -- Section~\ref{section.introduction}), and the actual cluster membership of some stars should be revisited.  
Table~\ref{table.mayrit} summarises my investigations (see references in the footnote).

%__________________________________________________ 
   \begin{table}
      \caption[]{Heliocentric distances to $\sigma$~Orionis$^{a}$.} 
         \label{table.distances}
     $$ 
         \begin{tabular}{cll}
            \hline
            \hline
            \noalign{\smallskip}
$d$ [pc] 			& Method				& Reference \\
            \noalign{\smallskip}
            \hline
            \noalign{\smallskip}
440				& Photometry			& Garrison 1967	\\          
400	  			& Dynamical parallax$^{b}$ & Heintz 1974	\\
370$\pm$50		& Spectroscopy$^{c}$	& Hunger et~al. 1989	\\
352$^{+166}_{-168}$ & Parallax			& Perryman et~al. 1997	\\
380$^{+136}_{-87}$ & Parallax			& Ma\'{\i}z-Apell\'aniz et al. 2004	\\
389$^{+34}_{-24}$	& Photometry			& Mayne \& Naylor 2008	\\
444$\pm$20		& Photometry$^{d}$		& Sherry et~al. 2008	\\
$\sim$385		& Dynamic parallax$^{e}$ & Caballero 20008b	\\
407$\pm$11		& Photometry			& Bell et~al. 2013	\\
380				& Interferometry		& Hummel et~al. 2013	\\
381$\pm$7		& Interferometry		& Schaefer 2013	\\
387.5$\pm$1.3		& Interferometry		& Schaefer et~al. 2016	\\
{360}$^{+44}_{-35}$	& Parallax$^{f}$		& {\em This work}	\\
\noalign{\smallskip}
            \hline
         \end{tabular}
     $$ 
\begin{list}{}{}
\item[$^{a}$] There are other determinations of the distance to the Ori~OB1b group by Brown et~al. (1994), de~Zeeuw et~al. (1999), and Hern\'andez et~al. (2005).
The parallactic determination by van~Leeuwen (2007), with a large uncertainty, is not tabulated either.
\item[$^{b}$] Under the double $\sigma$~Ori~AB scenario.
\item[$^{c}$] From a $\sigma$~Ori~D spectrogram.
\item[$^{d}$] For solar metallicity (Gonz\'alez-Hern\'andez et~al. 2008).
\item[$^{e}$] Under the triple $\sigma$~Ori~Aa,Ab,B scenario.
\item[$^{f}$] {Based on eight TGAS stars.}
\end{list}
   \end{table}

As shown in Table~\ref{table.brightest}, four of the {13} stars have parallaxes that deviate from the rest of measurements.
Of them, three have questionable features of youth or none at all:
($i$) Mayrit~(717307) = \object{TYC 4771--950--1} is an F7 star with very faint X-ray emission found only by Franciosini et~al. (2006) and very weak lithium absorption found by Caballero (2006).  
Although the Li~{\sc i} $\lambda\lambda$6707.76,6707.91\,{\AA} doublet starts to appear at the F/G spectral-type boundary in dwarf stars, it could be mistaken with the Fe~{\sc i} $\lambda$6707.43\,{\AA} line at low spectral resolutions;
($ii$) Mayrit~(1456284) =  \object{TYC 4770--1261--1} was detected only in one EPIC/{\em XMM-Newton} camera and with a low maximum likelihood parameter, so L\'opez-Santiago \& Caballero (2008) already called its true emission into question;
and ($iii$) Mayrit~(1227243) =  \object{HD 294275} is an early A-type dwarf with any signpost of youth.
The fourth star with deviating parallax is Mayrit~(459340) = \object{StHA~50}, an A2--6\,V star with a powerful H$\alpha$ emission (Caballero et~al. 2008) and a debris/evolved disc (Hern\'andez et~al. 2014).
Instead of being a UX~Ori-type star with a scattering edge-on disc in $\sigma$~Orionis, as proposed by Caballero et~al. (2008), it could be {an} unusual, isolated, Herbig Ae/Be star in the background, at about 300\,pc below the Galactic plane.
The Mayrit designations of the {four} stars are written in parenthesis in Tables~\ref{table.mayrit} and~\ref{table.brightest} and this paragraph for showing that they should not be used any more, as the stars are not actual $\sigma$~Orionis members.
A fifth star, also with a Mayrit designation in parenthesis and with only one controvertible signpost of youth (``fast rotation?'' -- Table~\ref{table.mayrit}), namely Mayrit~(1468100) = \object{HD~294301}, has an abnormally large total proper motion of 9.66\,mas\,a$^{-1}$. 
It could be either a young intermediate-mass star in the process of ejection via dynamical interactions within the Ori~OB1b association or, more likely, an old spectroscopic-binary interloper.

The other {eight} TGAS stars\footnote{According to Kharchenko et~al. (2004), four of the eight Mayrit/TGAS stars also had kinematic and photometric probabilities consistent with cluster membership at 2$\sigma$  ($p_{\rm kin}$, $p_{\rm ph} >$ 0.14).} towards the $\sigma$~Orionis cluster have conclusive youth features (OB spectral types, circumstellar discs, lithium in absorption, H$\alpha$ and/or X-rays in strong emission, flaring activity) and parallaxes that vary within a narrow range from 2.43$\pm$0.27\,mas (Mayrit~783254) to 3.24$\pm$0.52\,mas (Mayrit~863116).
{The mean parallax and standard deviation of the {eight} TGAS stars are mean$_8$($\pi$) = 2.78\,mas and std$_8$($\pi$) = 0.22\,mas, from which one can estimate a robust heliocentric distance.

However, alternative parallax values can be derived.
On the one hand, if the two most extreme cases (Mayrit~783254 and Mayrit~863116) are discarded, the mean parallax with {six} stars remains very similar but the standard deviations gets lower, with mean$_6$($\pi$) = 2.77\,mas and std$_6$($\pi$) = 0.06\,mas.
On the other case, if} three of the four independent HIP1 parallaxes are added to the set (it is not clear whether HD~37699 actually belongs to $\sigma$~Orionis, or to a new young population in the background, as StHA~50; Caballero 2007), the mean and standard deviation of the parallax with {11} TGAS+HIP1 stars are much the same: mean$_{11}$($\pi$) = 2.74\,mas and std$_{11}$($\pi$) = 0.21\,mas.
Weighted means, medians, Bayesian statistics, or even Lutz-Kelker corrections will {be worth investigating} only when the second {\em Gaia} data release is available in {April 2018}. 

Following the comment on correlated systematic terms and cluster mean parallaxes by {\em Gaia} Collaboration et~al. (2016), {by which a systematic component of
$\sim$0.3\,mas should be added to the parallax uncertainties}, the best estimate that I~can make with TGAS data for the $\sigma$~Orionis cluster is {2.78}$\pm$0.30\,mas, corresponding to a distance of {360}$^{+44}_{-35}$\,pc.

\section{Discussion and conclusions}
\label{section.discussion+conclusions}

The derived TGAS distance to $\sigma$~Orionis of $d$ = {360}$^{+44}_{-35}$\,pc fits in the lower end of all previous determinations, as comprehensively shown in Table~\ref{table.distances}.
In particular, the TGAS distance is equal within uncertainties to the most precise and accurate distance to $\sigma$~Orionis to date, by Schaefer et~al. (2016), who used an exhaustive set of interferometric and radial-velocity data.
We will have to wait for {\em Gaia} DR2 for a proper comparison and identification of systematic differences between both astrometric (from {space}) and interferometric+radial-velocity (from the ground) distances. 
In any case, this work paves the way for such a future comparison.

The TGAS distance resembles the values used by Zapatero-Osorio et~al. (2000), B\'ejar et~al. (1999, 2001), Lodieu et~al. (2009), and Pe\~na-Ram\'irez et~al. (2012) from Perryman et~al. (1997; {$d$ = 352$^{+166}_{-168}$}), and Caballero et~al. (2007) and Bihain et~al. (2009) from Brown et~al. (1994; $d$ = 360$^{+70}_{-60}$), which are the most representative works on the bottom of the initial mass function in the cluster.
Therefore, their results are not affected by the new TGAS distance (and only slightly by Schaefer et~al. 2016's).
However, the deprecated values of 400--450\,pc have been {used} in the last years by many teams worldwide for a variety of science topics, which may affect their results (e.g., Cody \& Hillenbrand 2011 for brown-dwarf variability; Williams et~al. 2013 for protoplanetary discs; Townsend et~al. 2013 and Oksala et~al. 2015 for the peculiar star $\sigma$~Ori~E; Koenig et~al. 2015 for deep infrared surveys). 

{The} mean proper motion of the {eight} Mayrit/TGAS cluster members is ($\mu_\alpha \cos{\delta}$, $\mu_\delta$) = (--0.2$\pm$1.5, +0.0$\pm$0.8)\,mas\,a$^{-1}$, which is consistent with previous determinations by Kharchenko et~al. (2005) and Caballero (2007,~2010).
{This low mean proper motion}, due to the location of $\sigma$~Orionis in the antapex and at almost 400\,pc, will not prevent that after further {\em Gaia} data releases the internal kinematic dispersion of the open cluster is investigated in a per-star basis (at the end of the {\em Gaia} mission, some $\sigma$~Orionis stars will have proper motion uncertainties of only a few tens $\mu$as\,a$^{-1}$). 
Proper-motion analysis has been already used for discarding photometric cluster member candidates in the cluster ({Kharchenko et~al. 2004}; Caballero 2007, 2010; Lodieu et~al. 2009), and it seems to be the most efficient rejecting method for A- and F-type stars, which signposts of youth are not as evident as in other spectral types (such as the Li~{\sc i} absorption doublet in fainter, very young GKM-type stars and brown dwarfs). 

Although the derived TGAS distance to $\sigma$~Orionis supersedes most previous determinations and is consistent with the value at 387.5$\pm$1.3\,pc of Schaefer et~al. (2016), which should be used as a reference until the analyses of {\em Gaia} DR2 data are made public, perhaps the most striking result in this work was obtained in the process of selection of bona~fide cluster members.
In this process, I~discarded {five} Mayrit stars from the initial list of targets (including the Herbig Ae/Be star StHA~50), which represent more than half of all previously considered cluster member candidates with masses between 1.2\,$M_\odot$ and 2.1\,$M_\odot$ (Caballero 2007).
In other words, there could be more B stars in $\sigma$~Orionis than A and F stars together.
Besides, the addition of a third massive star in the central astrometric ``binary'', $\sigma$~Ori~Ab ($\mathcal{M} \sim$ 12.8\,$M_\odot$ -- Sim\'on-D\'iaz et~al. 2015; Schaefer et~al. 2016), {leads to a flattening of} the slope of the mass spectrum with respect to previous determinations at the high-mass end (Caballero 2007; Pe\~na-Ram\'irez et~al. 2012).
The slope value can be at about $\alpha$ ($\equiv -\gamma \equiv 1-\Gamma$) $\approx$ +1.5 in the stellar mass interval from 1.2 to 17\,$M_\odot$, well below the Salpeter's one at +2.35.
A careful determination of the actual slopes of the $\sigma$~Orionis mass spectrum with only Tycho-2 data ({using roughly the same star sample as is in this work with TGAS data}) was already shown to be inconclusive by Caballero (2011), while the use of the full {\em Gaia}~DR1 dataset is out of the scope of this work. 

With the same methodology as the one presented here, and when new precise parallaxes, proper motions, and $B_P$ and $R_P$ photometry from {\em Gaia}~DR2 are available for a much larger number of stars, it will be possible to go one step ahead, and create a clean list of bona~fide $\sigma$~Orionis members from the high-mass end down to the star/brown boundary, {which will help measuring} more accurate cluster mass function indices, isochronal ages, and disc and activity fractions.

%------------------------------------------------------------------------------

\acknowledgements
{JAC thanks the referee, R.-D. Scholz, for his constructive input.}
%JAC {was} a Klaus Tschira Stiftung fellow at the Landessternwarte K\"onigstuhl;
%{he is now a {\em joven investigador} of the CSIC at the Centro de Astrobiolog\'{\i}a.}
Financial support was provided by the Klaus Tschira Stiftung and Spanish Ministerio de Econom\'{\i}a y Competitividad under grants AYA2014-54348-C3-2-R {and AYA2015-74151-JIN}.
This research made use of the SIMBAD, operated at Centre de Donn\'ees astronomiques de Strasbourg, France, the NASA's Astrophysics Data System, and of data from the European Space Agency (ESA) mission {\it Gaia} ({\tt http://www.cosmos.esa.int/gaia}), processed by the {\it Gaia} Data Processing and Analysis Consortium (DPAC, {\tt http://www.cosmos.esa.int/web/gaia/dpac/consortium}).
Funding for the DPAC has been provided by national institutions, in particular the institutions participating in the {\it Gaia} Multilateral Agreement.

%------------------------------------------------------------------------------

%\bibliographystyle{an}
%\bibliography{an-demo}
% 

%------------------------------------------------------------------------------

\appendix

\section{Table~\ref{table.brightest}}

%__________________________________________________ 
   \begin{table*}
      \caption[]{Relevant astro-photometric data of the brightest stars in the $\sigma$~Orionis cluster$^{a}$.} 
         \label{table.brightest}
     $$ 
         \begin{tabular}{lll ccc cc l}
            \hline
            \hline
            \noalign{\smallskip}
Name  			& Mayrit 		&Sp. 		& $V_T$	& $G$	& $J$	& $\mu$			& $\pi$			& Ref.		\\
				& 			& type		& [mag]	& [mag]	& [mag]	& [mas\,a$^{-1}$]	& [mas]			&			\\
            \noalign{\smallskip}
            \hline
            \noalign{\smallskip}
$\sigma$ Ori Aa,Ab,B	& AB		& O9.5\,V+	&  3.763& ...   &  4.752		&  4.63			& 2.84 $\pm$ 0.91	& HIP1		\\	       
$\sigma$ Ori C	    	& 11238   		& A2\,V          	&  9.43$^b$ &  9.450&  9.086	&  0.00			& ...		       		& NOMAD		\\	     
$\sigma$ Ori D	    	& 13084   		& B2\,V          	&  6.557& ...   &  7.116		&  4.62			& (2.84 $\pm$ 0.91)	& HIP1		\\	     
$\sigma$ Ori E	    	& 42062   		& B2\,Vp         	&  6.344& ...   &  6.974		&  2.61			& ...		       		& TYC1		\\	     
HD 294272 A	    	& 182305$^{c}$& B9.5\,III      	&  8.389&  8.464&  8.346		&  2.07	      		& 2.75 $\pm$ 0.34    	& {TGAS}  	\\ 
HD 294272 B	    	& 189303$^{c}$& B8\,V          	&  8.554&  8.766&  8.779		&  0.67	      		& 2.75 $\pm$ 0.42	& {TGAS}  	\\ 
HD 294271	    	& 208324  	& B5\,V          	&  7.856& ...   &  8.100		&  0.72			& ...		       		& TYC2		\\	     
HD 37525 AB	    	& 306125  	& B5\,Vp         	&  8.058&  8.088&  8.131		&  2.14			& 2.55 $\pm$ 0.99	& HIP1       	\\	    
StHA 50 	    		& (459340)   	& A2--6\,Ve      	& 11.275& 11.180& 10.666	&  4.25	  		& 0.76 $\pm$ 0.31  	& {TGAS}  	\\ 
HD 294273	    	& 521210  	& A3          		& 10.828& 10.606& 10.176	&  4.64			& ...     			& TYC1		\\  	    
HD 37564	    		& 524060$^{c}$& A8:\,V         	&  8.443&  8.427&  7.976		&  0.35	  		& 2.85 $\pm$ 0.36   	& {TGAS}   	\\ 
TYC 4771-950-1     	& (717307)   	& F7          		& 11.429& 11.084& 10.088	& 14.62	  		& 1.90 $\pm$ 0.27   	& {TGAS}  	\\ 
TYC 4771-661-1      	& ...     		& K-M:        	& 11.899& 11.641& 10.617	& 26.07			& ...     			& TYC2		\\  	    
TYC 4771-720-1      	& ...     		& K-M:        	& 11.787& 11.331& 10.269	& 26.07			& ...     			& TYC2		\\  	    
2E 1455 	    		& 783254$^{c}$& ...	        		& 11.015& 10.670&  9.255	&  0.56	  		& 2.43 $\pm$ 0.27   	& {TGAS}  	\\ 
HD 294278	    	& ...     		& K2           	&  9.941&  9.310&  7.592		&  8.54	  		& 2.25 $\pm$ 0.28    & {TGAS}  		\\ 
HD 294300 AB	    	& 863116$^{c}$& G5-K0       	& 10.195&  9.657&  8.462		&  0.88	  		& 3.24 $\pm$ 0.52   	& {TGAS}  	\\ 
HD 294269	    	& ...     		& G0          	& 10.795& 10.379&  9.178	& 44.28	  		& 4.12 $\pm$ 0.23   	& {TGAS}  	\\ 
V1147 Ori	    		& 960106$^{c}$& B9\,IIIp       	&  8.992&  9.026&  8.876		&  1.97	  		& 2.77 $\pm$ 0.35   	& {TGAS}   	\\ 
TYC 4771-962-1      	& 968292		& ...	        		& 11.101& 11.000& 10.202	&  3.89			& ...       			& TYC2		\\	      
TYC 4771-934-1      	& ...     		& K-M:        	& 12.870& ...	&  8.948		&  7.97	 		& ...       			& TYC2		\\	      
HD 294274	    	& ...     		& G0          	& 10.758& 10.497&  9.428	& 21.39	  		& 3.06 $\pm$ 0.25   	& {TGAS}  	\\ 
TYC 4771-621-1   	& ...     		& K-M:        	& 10.920& 10.642&  9.796	& 13.12			& ...       			& TYC2		\\	      
TYC 4771-873-1	& ...	   		& F7-9        	& 12.031& 12.097& 11.125	&  6.12	  		& 1.22 $\pm$ 0.57   	& {TGAS}  	\\ 
HD 37333	    		& 1116300 	& A1\,Va         	&  8.508&  8.529&  8.413		&  5.11			& 2.52 $\pm$ 1.08	& HIP1       	\\  	    
HD 294279	    	& 1160190$^{c}$& F3-5        	& 10.685& 10.587&  9.873	&  3.06	 		& 2.80 $\pm$ 0.25   	& {TGAS} 	 	\\ 
TYC 4771-1468-1     & ...     		& K-M:        	& 11.136& 10.466&  8.616	&  5.73	  		& 0.53 $\pm$ 0.27   	& {TGAS}  	\\ 
HD 294275	    	& (1227243)   	& A1\,V          	&  9.431&  9.419&  9.256		&  3.50	  		& 3.53 $\pm$ 0.36    & {TGAS}  		\\ 
HD 294277	    	& ...     		& K2           	&  9.737&  8.810&  6.773		&  8.24	  		& 1.57 $\pm$ 0.30    & {TGAS}  		\\ 
HD 294268	    	& 1285339$^{c}$& F5         	& 10.507& 10.262&  9.393	&  1.01	  		& 2.68 $\pm$ 0.47   	& {TGAS}  	\\ 
HD 37545	    		& 1288163 	& B9\,V          	&  9.277&  9.306&  9.261		&  2.20			& ...     			& TYC2		\\  	    
HD 294270	    	& ...     		& G0          	& 10.960& 10.665&  9.780	& 28.81			& ...       			& TYC2		\\	      
TYC 4770-924-1      	& ...     		& K-M:        	& 11.938& 11.947 & 10.725	& 36.08	  		& 5.19 $\pm$ 0.82	& {TGAS}  	\\ % G = 11.947
HD 37686	    		& 1359077 	& B9.5\,Vn       	&  9.186& ...   &  9.207		&  2.97			& ...      			& TYC2		\\	      
HD 294298	    	& 1366055 	& G0:         	& 10.975& 10.608&  9.339	&  1.84			& ...       			& TYC2		\\	      
TYC 4770-1432-1     & ...     		& K-M:        	& 12.337& 12.158   & 11.077	& 30.90	  		& ...	      			& TYC2		\\ % G = 11.686	      
TYC 4770-1261-1     & (1456284)	&  ...     		& 11.498& 11.686& 10.721	&  9.53	  		& 3.94 $\pm$ 0.85   	& {TGAS}  	\\ 
HD 294301	    	& 1468100 	& F2\,V(n)      	& 11.139& 10.905& 10.210	&  9.66	  		& 2.64 $\pm$ 0.24   	& {TGAS} 		\\ 
HD 294276	    	& ...     		& G0          	& 10.418& 10.203&  9.183	& 62.32			& ...       			& TYC1		\\	      
HD 37699	    		& 1548068 	& B5\,Vn         	&  7.590& ...      &  7.841		&  0.95			& 0.63 $\pm$ 1.01	& HIP1	 	\\	       
TYC 4771-1012-1    	&  ...     		& K-M:        	& 11.263& 10.865&  8.998	&  9.74			& ...      			& TYC2		\\ 	     
HD 294297	    	& ...     		& F6-8        	& 10.193&  9.925&  9.062		& 30.59			& ...      			& TYC1		\\ 	     
HD 294307	    	& ...     		& F8          		& 10.552& 10.256&  9.439	& 21.61	  		& 4.38 $\pm$ 0.28   	& {TGAS}  	\\ 
TYC 4770-1129-1    	& ...			& ...	        		& 11.827& 12.171& 11.232	&  5.67	  		& 0.91 $\pm$ 0.38   	& {TGAS} 		\\ 
HD 294280	   	& ...     		& K5           	&  9.865&  8.774&  6.429		&  4.03	  		& 1.68 $\pm$ 0.33    	& {TGAS}  	\\ 
TYC 4770-1018-1    	& ...     		& K-M:        	& 11.224& 10.426&  8.620	&  5.02	  		& 0.93 $\pm$ 0.25   	& {TGAS}  	\\         
\noalign{\smallskip}
            \hline
         \end{tabular}
     $$ 
\begin{list}{}{}
\item[$^{a}$] {Star name, Mayrit designation, spectral type, Tycho-2 $V_T$, {\em Gaia} $G$, and 2MASS $J$ magnitudes, total proper motion, parallax, and reference for the last two parameters.
{Stars are sorted by angular separation to $\sigma$~Ori~Aa,Ab,B.}
{\em An extended version of this table, with coordinates, Tycho-2, and TGAS proper motions, and magnitudes from $B_T$ to $W4$, is (WILL BE) available online via VizieR}.}
\item[$^{b}$] {$V_T$ magnitude of $\sigma$~Ori~C derived from $V$ in Sherry et~al. (2008).}
\item[$^{c}$] {The {eight} Mayrit/TGAS stars used for the parallax determination.}
\end{list}
   \end{table*}
\end{document}